# Assessing the varying level of impact measurement accuracy as a function of the citation window length[1]

*Giovanni Abramo*[a,b,*], *Tindaro Cicero*[b], *Ciriaco Andrea D'Angelo*[b]

[a] Institute for System Analysis and Computer Science (IASI-CNR)
National Research Council of Italy

[b] Laboratory for Studies of Research and Technology Transfer
School of Engineering, Department of Management
University of Rome "Tor Vergata"

**Abstract**

With the passage of more time from the original date of publication, the measure of the impact of scientific works using subsequent citation counts becomes more accurate. However the measurement of individual and organizational research productivity should ideally refer to a period with closing date just prior to the evaluation exercise. Therefore it is necessary to compromise between accuracy and timeliness. This work attempts to provide an order of magnitude for the error in measurement that occurs with decreasing the time lapse between date of publication and citation count. The analysis is conducted by scientific discipline on the basis of publications indexed in the Thomson Reuters Italian National Citation Report.

**\* Corresponding author:** Dipartimento di Ingegneria dell'Impresa, Università degli Studi di Roma "Tor Vergata", Via del Politecnico 1, 00133 Rome - ITALY, tel. +39 06 72597362, giovanni.abramo@uniroma2.it

# 1. Introduction

The use of national exercises to evaluate research systems is becoming ever more diffuse. One of the major objectives is to support efficient allocation of public resources to the various actors in the national systems. Traditionally the assessment exercises relied on peer review approaches, but advances in bibliometric techniques have led to many governments adopting bibliometric indicators to inform or even entirely substitute peer review, at least for the hard sciences. The penetration of bibliometrics can be appreciated by examining the typologies of three assessment frameworks: the Research Excellence Framework (REF) in the UK, the Quinquennial Research Evaluation (VQR) in Italy, and the Excellence in Research for Australia initiative (ERA). For the ERA, preparations for submissions began in June 2010. For the VQR, detailed guidance on submissions and assessment criteria is expected in 2011. For the United Kingdom REF, guidelines will be published during 2011, with institutions invited to make submissions during 2013 and actual assessment taking place in 2014. The REF is a typical example of a so called "informed peer-review" exercise, where the assessment outcomes will be a product of expert review informed by citation information and other quantitative indicators. It will substitute the previous Research Assessment Exercise series which were pure peer-review. The Italian VQR, substituting the previous pure peer-review Triennial Evaluation Exercise (VTR), can be considered a hybrid: a varying mix of pure peer-review, informed peer-review and the bibliometric approach. To prepare judgments of research output quality, the panels of experts appointed in each of fourteen disciplines, can choose one or both of two methodologies for evaluating any particular output: i) citation analysis; and/or ii) peer-review by external experts, selected by a collegial decision of the panel. The Australian ERA assessment in the hard sciences is conducted through a pure bibliometric approach. Single research outputs are evaluated by a citation index referring to world and Australian benchmarks. Because the entire research staff of the institutions must submit their full research product, indicators of research volume are also used to evaluate overall research performance.

Studies have demonstrated that there is indeed a positive relationship between citations of a work and the opinions of experts concerning its quality (Oppenheim, 1997; Rinia et al., 1998; Aksnes and Taxt, 2004; Reale et al., 2006; Franceschet and Costantini, 2011), however there are numerous differences between peer review and bibliometrics, including in the limitations of the two approaches.

Peer review presents a series of well documented and much discussed limitations regarding each of its three fundamental steps: i) the choice of products for submission to evaluation; ii) the choice of experts entrusted with evaluation of the products; iii) the inherent subjectivity in the judgments given by the reviewer, as offered for each product (Moxham and Anderson, 1992; Horrobin, 1990, Bornmann 2008). However bibliometrics also has its own limitations. The most notable is the fact that it can only be applied to disciplines where publication in journals is considered a reliable proxy of research output, meaning only the hard sciences (Moed, 2005). For the hard sciences, Abramo and D'Angelo (2011) have compared the results of the Italian VTR with those from a bibliometric simulation and have shown that bibliometric approach is greatly preferable to peer review for accuracy, robustness and functionality of measurement, and for the costs and times involved. However the 2011 study by Abramo and D'Angelo did not deal with the critical concern of the time that elapses between date of publication and the date of actually counting the citations, which is necessary to obtain citation counts that can give an accurate measure of the true publication impact. In theory the peer-review approach would permit an evaluation of quality immediately on release of the publication, but with bibliometrics the citations of a work can only be a good proxy of true impact when there has been a sufficient lapse from the date of publication. A minimum "citation window" is necessary. The potential problem is that whatever the intentions for the evaluation exercise (selective funding, informing research policies and management decisions, reducing information asymmetry between suppliers of knowledge and users), it is highly desirable that the evaluation results be available in close reference to the period being evaluated. This factor could affect the applicability of bibliometric methods. We note that in spite of the questions raised, the time necessary to implement peer-review exercises is longer (two years or more for the entirety of steps) than for the mechanisms of bibliometric exercise. Also, peer-review exercises typically occur over cycles of 5 to 6 years (the latest RAE covered an eight year period), which is slower and less frequent than is desirable for evaluation aims. Evaluations based on bibliometric techniques can be more frequent and thus more effective in stimulating continuous improvement in the research system.

Given the concerns, it is of great interest to understand the number of years necessary before citations of a publication can be considered an accurate and robust proxy of real scientific impact, and if this window of time differs from one discipline to another. The present work intends to provide answers to these questions and to define a methodology (in terms of citation windows) for conduct of bibliometric exercises that will offer the necessary robust ratings and rankings.

There are not many works that have dealt with this question. In general, we can say that citations have increased gradually over time, as shown by the growing value of journals' impact factors; moreover, impact factors vary widely across fields (Althouse et al., 2008). Glanzel et al. (2003), analyzing a set of works published in 1980 and indexed in the Science Citation Index (SCI), demonstrate that the probability of publications that are not cited or poorly cited over an initial period of 3 to 5 years should then become highly cited beyond the standard bibliometric time horizon (i.e. in a time window of 21 years after publication) is very remote and limited to rare exceptions. However, like Rousseau (1988), they note that in certain fields (e.g. mathematics-related), the standard bibliometric time horizon is greater than in others: for correct evaluation of impact of a work in mathematics the citation window should be more than three years. A subsequent study by Adams (2005) includes a conclusion that "initial citation counts" (i.e. citations received 1 and 2 years after publication) "might be useful as a forward indicator of the long-term quality of research publications". This author's findings are based on observation of publications for 1993 in the life and physical sciences, extracted from the UK National Citation Report licensed from Thomson Reuters by the UK Office of Science and Technology (OST). Considering a window from 1993 to 2002, Adams detects a strong correlation between the ranking lists for publications per number of citations in the first two years and continuing over subsequent years. Stringer et al. (2008), investigate the time scale for the full impact of papers published in a given journal to become apparent, and find that it varies from less than 1 year to 26 years, depending on the journal. Continuing from this previous literature, the intention of our current work is to study how accuracy in the measurement of publication impact varies, in each hard science discipline, in function of the length of the citation window between date of publication and citation count. All the limits of citation counts as proxy of impact, amply discussed in the literature (MacRoberts and MacRoberts, 1989; Moed, 2005; Glanzel, 2008), remain. Furthermore, we also report on the issues of the citation patterns seen in the various subject categories, the first-citation speed, and the error in evaluating a publication as having nil impact when it has not matured any citations within a given date. The study is based on publications indexed in the Thomson Reuters 2001-2008 Italian National Citation Report, extracted from Web of Science (WoS).

The following section describes the dataset used for the analysis, the elaborations, and the results concerning the accuracy of measure of impact in function of the length of the citation window. The final section provides a summary of the main findings, discusses their implications, and indicates opportunities for further consideration and examination.

## 3. Results and analysis

The dataset used for the analysis was extracted from the Italian Observatory of Public Research (ORP[2]), a database developed by the authors and derived under license from Thomson Reuters Italian National Citation Report. The ORP contains all scientific publications involving an author from an Italian research organization (95 universities, 76 research institutions and 192 hospitals and health care research organizations). The data for the analysis refer to the production for 2001, which is a total of 37,430 scientific publications[3]. The field of observation is limited to the 171 WoS subject categories of the hard sciences, grouped into 8 disciplines[4]: Biology, Biomedical research, Chemistry, Clinical medicine, Earth and space science, Engineering, Mathematics and Physics. The citations are observed as of December 31 of each year, from 2001 to 2008.

The following subsections provide the results of the analysis. The first part gives a descriptive analysis highlighting the differences between the disciplines in terms of citations patterns. The second part characterizes the function of "first-citation speed", meaning the probability function for when the first citation will be received in the years succeeding the publication of a work. The last subsection addresses the specific research question that inspires the work, which is the query of how many years must lapse between publication and observation before the citations of a work can be considered an accurate proxy of its real impact on scientific progress.

### 3.1 Citation profiles per discipline

The literature has long noted that among the different scientific fields there are differences in citation patterns and in the speed that publications accumulate citations (Garfield, 1972; Egghe and Rousseau, 1990; Moed, 2005, Althouse et al. 2009). This is due to a number of factors: i) different numbers of journals indexed for the fields in the main bibliometric databases, such as Web of Science or Scopus; ii) different citation practices among fields; and least, but not last iii) different production functions across fields. In this section we attempt to characterize this citation variability trough analysis, by field, of the average value of citations received for publications in the dataset. Figure 1 provides an example of the analysis, showing the average value of citations received per year for the 1,819 Italian publications for 2001 in the WoS subject category Astronomy and astrophysics. The curve shows a maximum for average value of citations (3.25) in 2003, followed by a decrease to an average of 1.6 citations in 2007 and then a further slight increase between 2007 and 2008.

---

[2] www.orp.researchvalue.it, last accessed on April 20, 2011.
[3] For further significance, the same analysis reported in this paper was repeated for publications from 2002: the results and findings were completely in accord with those presented here.
[4] The discipline classification of the subject categories refers to a past ISI classification which no longer exists. Its plausibility may be checked at
http://www.disp.uniroma2.it/laboratoriortt/TESTI/Altro/WoS_classif.pdf

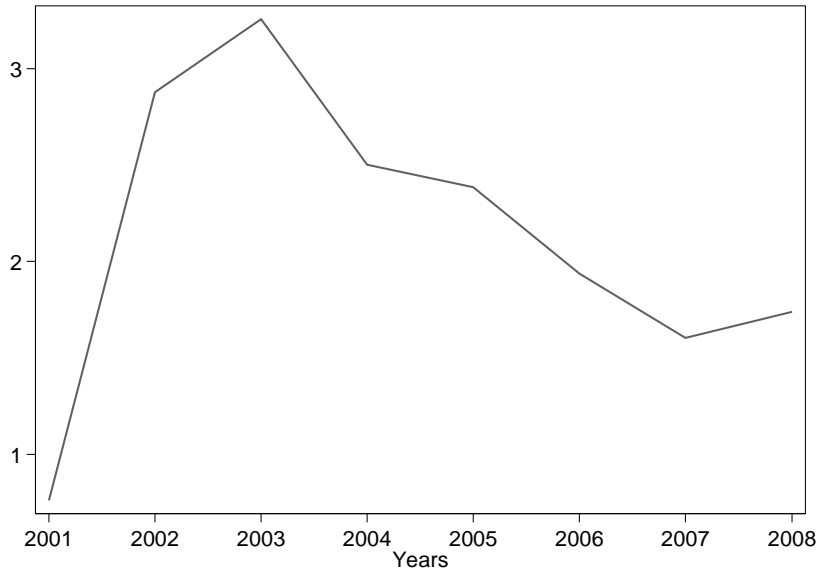

*Figure 1: Average annual citations for Italian publications in Astronomy and astrophysics (1,818) for 2001*

To detect differences between scientific fields, the publications were grouped by discipline and the analysis for "Astronomy and astrophysics" was repeated for each discipline. This exercise showed three clusters characterized by different citation patterns. The first cluster identified is the group of Biology, Chemistry and Clinical medicine (Figure 2). The graph shows a bimodal distribution with peaks at two and four years from publication. A third and lesser peak is seen in the final year observed (2008). Among these disciplines, Clinical medicine registers the highest average maximums. Publications in this discipline received an average of 2.81 citations in 2003 and 2.82 in 2005. Average values for these peak years are lower in Biology (2.48 and 2.44) and Chemistry (2.03 and 2.10).

The second cluster consists of Biomedical research and Physics. Here the trend observed is generally a decrease after 2003: like for the first cluster, this is the peak year for citations. Biomedical research actually shows three gradually descending peaks in citations, in the second, fourth and sixth year from publication. The absolute levels of average citations resemble those for the first cluster of disciplines. In Physics, the oscillations are less pronounced after the peak of 2.01 average citations in 2003.

The third cluster (Engineering, Earth and space science and Mathematics) is evidently different than the other disciplines: there is a trend to increasing citations throughout the citation window, with a peak in the final year of 2008 (Figure 3). The average intensity of citations is also less than for the other disciplines.

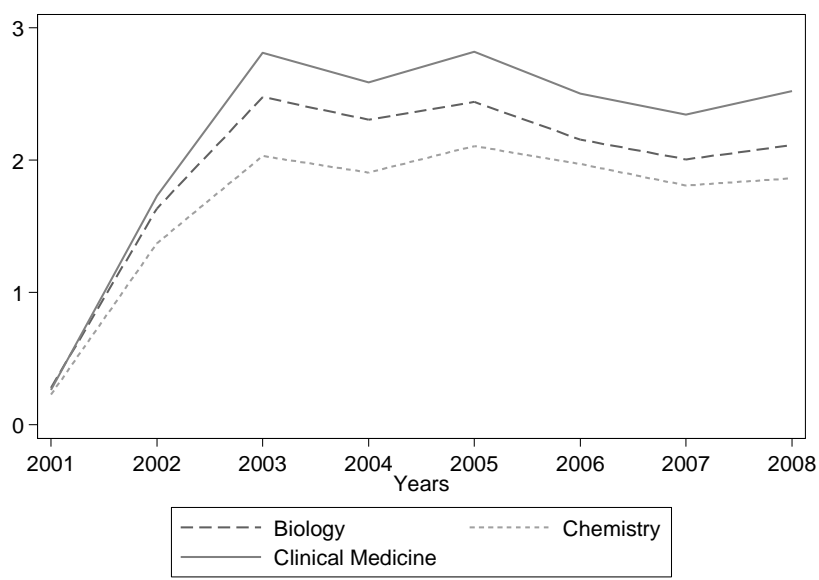

*Figure 2: Average annual citations for publications in Biology (6,041), Chemistry (4,191) and Clinical medicine (9,418) for 2001*

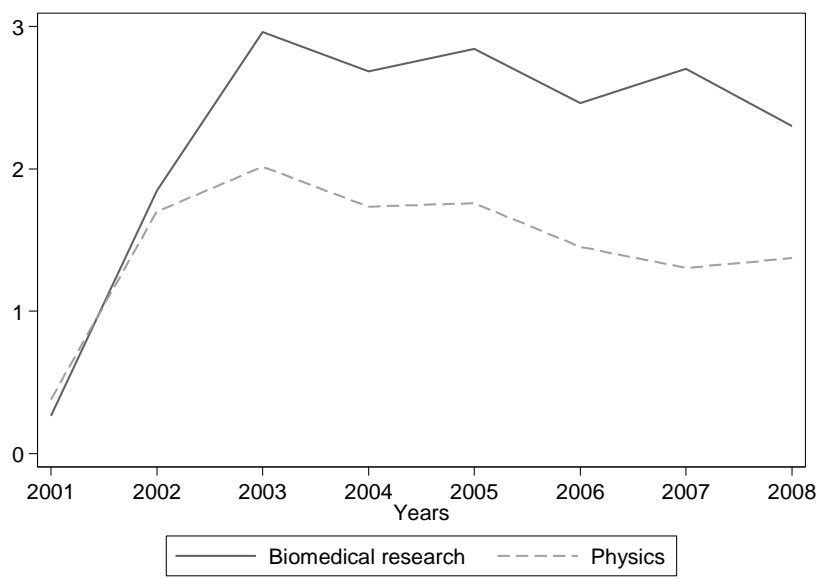

*Figure 3: Average annual citations for publications in Biomedical research (6,108) and Physics (8,901) for 2001*

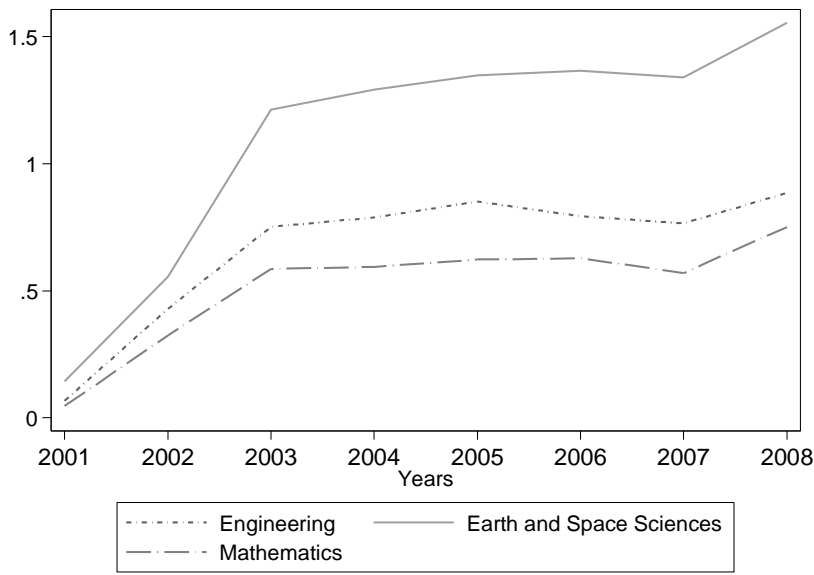

*Figure 4: Average annual citations for Italian publications in Engineering (8,062), Earth and space science (2,436) and Mathematics (2,008) for 2001*

The progressions illustrated are in only partial agreement with observations by Moed (2005): comparing time distribution of citations for the two research fields of biochemistry-molecular biology and mathematics, he observed that in the first field the peak of the distribution is at an average of two years from publication, while for the second the peak in citations was reached an average of a year later.

It is also interesting to know if the progress of citations over time depends on the intrinsic quality of the individual publication: in other words, if publications that are immediately highly cited continue with citation patterns that are different than those for publications that are initially poorly cited.

To establish this we again refer to the publications for Astronomy and astrophysics, for the period graphed in Figure 1. We identify two subsets of publications: the first is those that have only one citation as of 31/12/2001 (the "poorly cited" subset); the second is a group that place in the top national decile as of the same date ("highly cited"). Figure 5 shows the progress of average citations for these two subsets. Once again we see the peak of 2003 (4.91 average citations for the poorly-cited and 13.46 for the highly-cited) after which the trend is clearly decreasing for both subsets. The trend for the highly cited publications is less constant, actually showing an interruption in 2008.

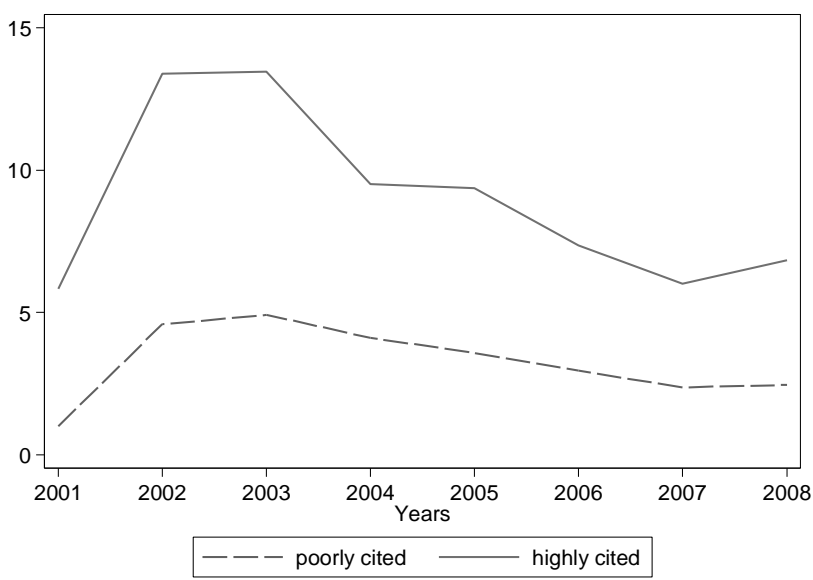

*Figure 5: Average annual citations for Italian publications in Astronomy and Astrophysics, subdivided by "poorly" (203) and "highly" cited (182), for 2001*

This analysis was repeated for the publications of every discipline, showing that the citation pattern of poorly cited publications is consistently very similar to that for highly cited publications.

**3.2 First-citation speed analysis**

In this section we analyze the time at which a publication receives its first citation. The scientific impact of publications is actually sometimes evaluated through "immediacy", or "first-citation speed" (Bormann, 2010; Van Dalen and Henkens, 2005).

The objective here is to evaluate how probability of receiving a first citation changes over time, and thus indicate the potential for error in considering real impact as nil, when observing works that have not received any citations after a series of years. Table 1 presents the case of the 6,041 publications in Biology. Of these, 77.8% (total 4,702) receive at least one citation over the eight years observed. But by December 31 of the actual publication year (2001), 879 works (14.6%) have received at least one citation. In the next year, the probability of receiving first citation rises to 36.2% of the total. By the third year after publication the probability of receiving a first citation begins to decrease: by 31/12/2003 the share that has received first citation is already 66.7%, compared to the final total of 77.8% observed at 31/12/2008. Column 4 of the table shows the evident and linear reduction in probability of receiving a first citation after two years have passed from publication.

The analysis was repeated for all disciplines. Table 2 presents a summary of results observed at December 31 of each year. At the end of the window of observation, the percentage of publications cited varies from 84.5% in Chemistry to 51.0% for Engineering. Of the publications cited, at least 18% had already received their first citation in the actual year of publication in Biology, Biomedical research, Chemistry, Clinical medicine, and for Physics the level reached almost 25%. Within the next year, these same disciplines show first-citation percentages ranging from 63.4% in Clinical medicine to 70.0% for Physics. A second cluster of disciplines shows slower speeds: by the close of 2002, 38.3% of works in Mathematics receive first citation, while the figures are 48.0% in Engineering and 50.6% in Earth and space science. At the close of 2003, over 85% of the publications in the first cluster have received first citation. Further back, but with the gap constantly decreasing, are the second group of disciplines: the constant overall slowing of first-cited publications is clear.

| Date | First-cited at the date | Ratio of total publications (6,041) | Cumulative of total publications | Ratio of total cited publications (4,702) | Cumulative of total cited publications |
|---|---|---|---|---|---|
| 31/12/2001 | 879 | 14.6% | 14.6% | 18.7% | 18.7% |
| 31/12/2002 | 2187 | 36.2% | 50.7% | 46.5% | 65.2% |
| 31/12/2003 | 962 | 15.9% | 66.7% | 20.5% | 85.7% |
| 31/12/2004 | 319 | 5.3% | 72.0% | 6.8% | 92.4% |
| 31/12/2005 | 164 | 2.7% | 74.7% | 3.5% | 95.9% |
| 31/12/2006 | 89 | 1.5% | 76.1% | 1.9% | 97.8% |
| 31/12/2007 | 56 | 0.9% | 77.1% | 1.2% | 99.0% |
| 31/12/2008 | 46 | 0.8% | 77.8% | 1.0% | 100.0% |

*Table 1: First-citation speed of Italian publications in Biology for 2001*

| Discipline | Cited at 31/12/2008 (% of total) | Cumulative (% of total cited publications) | | | | | | |
|---|---|---|---|---|---|---|---|---|
| | | 2001 | 2002 | 2003 | 2004 | 2005 | 2006 | 2007 |
| Biology | 77.8 | 18.8 | 65.3 | 85.7 | 92.5 | 96.0 | 97.8 | 99.1 |
| Biomedical research | 76.1 | 18.7 | 66.5 | 87.6 | 93.7 | 97.0 | 98.4 | 99.3 |
| Chemistry | 84.5 | 18.7 | 67.6 | 87.6 | 93.4 | 96.6 | 98.2 | 99.3 |
| Clinical medicine | 78.2 | 17.9 | 63.4 | 85.4 | 92.5 | 96.3 | 98.0 | 99.1 |
| Earth and space science | 62.2 | 14.6 | 50.6 | 77.7 | 89.2 | 93.7 | 96.5 | 98.9 |
| Engineering | 51.0 | 10.4 | 48.0 | 73.7 | 84.9 | 91.6 | 94.5 | 97.6 |
| Mathematics | 55.4 | 7.4 | 38.3 | 64.8 | 78.5 | 86.6 | 93.5 | 96.6 |
| Physics | 67.4 | 24.8 | 70.0 | 86.2 | 92.4 | 96.0 | 97.8 | 99.1 |

*Table 2: Time series of cumulative first-cited publications on total cited 2001 Italian publications*

### 3.3 Accuracy of impact measurement as a function of the citation window length

In this section we analyze variation in the accuracy of impact measure in function of the time lapse between date of publication and citation count. As reference for maximum accuracy we take the cumulative citations for the publications observed as of 31/12/2008: effectively we assume the cumulative window of eight years as offering maximum accuracy. Even though citations will continue to accumulate after eight years we can certainly consider this assumed time period as being well over what a decision maker would desire, in terms of obtaining information useful for policy and management decisions from an evaluation. We also note that the intention of assessment exercises is not to so much to measure the absolute impact of research products as to understand their relative impact, or in other words to know whether the various products have greater or lesser impact than others.

For each subject category and every year, beginning 2001, we rank the publications by cumulative citations and conduct a Spearman correlation analysis for these annual rankings relative to the 2008 benchmark. Table 3 shows the example of results for only a single field within each discipline, the largest in terms of number of publications. The table shows that the correlation is weak for the 2001 observations in all fields, with the exception of Astronomy and astrophysics. From the next year (31/12/2002) onward, the correlation becomes stronger and stronger. The maximum 2002 correlation is once again seen for Astronomy and astrophysics (0.910), followed by Biochemistry and molecular biology (0.823), Oncology (0.815) and Neurosciences (0.806). The lowest correlation, as would be expected, is in Mathematics (Mathematics, applied) at 0.593. However even in this field, by the next year (at 31/12/2003), correlation with the benchmark rises to 0.793, and for all the other fields becomes very strong, always greater than 0.85.

| Subject category (total publications) | 2001 | 2002 | 2003 | 2004 | 2005 | 2006 | 2007 |
|---|---|---|---|---|---|---|---|
| Biochemistry & molecular biology (2,018) | 0.420 | 0.823 | 0.931 | 0.964 | 0.981 | 0.991 | 0.997 |
| Oncology (1,631) | 0.439 | 0.815 | 0.933 | 0.965 | 0.982 | 0.991 | 0.997 |
| Chemistry, physical (1,203) | 0.282 | 0.694 | 0.871 | 0.932 | 0.965 | 0.983 | 0.994 |
| Neurosciences (1,543) | 0.400 | 0.806 | 0.932 | 0.968 | 0.985 | 0.992 | 0.998 |
| Environmental sciences (883) | 0.332 | 0.669 | 0.877 | 0.932 | 0.966 | 0.983 | 0.994 |
| Engineering, electrical-electronic (2,342) | 0.244 | 0.673 | 0.850 | 0.921 | 0.962 | 0.976 | 0.990 |
| Mathematics, applied (926) | 0.264 | 0.593 | 0.793 | 0.887 | 0.931 | 0.972 | 0.956 |
| Astronomy and astrophysics (1,818) | 0.595 | 0.910 | 0.965 | 0.980 | 0.991 | 0.995 | 0.998 |

*Table 3: Spearman correlation between rankings of publications based on cumulated citations at each year and 2008 ranking*

In most real-world assessment exercises the publications are grouped in classes, according to their impact. Similarly, here we have classified the publications into five sets. The first group is "nil impact", consisting of those that, year after year, receive no citations. For the other publications we assign impact values of 4, 3, 2 and 1, corresponding to the first, second, third and fourth quartiles for the distributions by subject category, for citations accumulated, in each successive year. For each subject category and each year we can thus measure the average difference in classification for the publications compared to the benchmark year of 2008[5]. Table 4 presents the average values of shift for the five classes, by discipline. It clearly illustrates that the differences are gradually less and less as the date nears the benchmark. The classifications for 2001 would be almost useless, with average differences in classification of more than one for all disciplines, with values as high as 1.6 for Clinical medicine and biology. From 2002 on, the greatest shifts in classification are seen in Mathematics, beginning from an average move of almost one place (0.925) and gradually descending to 0.171 by 31/12/2007. The error is notably less for all the other disciplines. Convergence arrives quickest in Physics, followed by Biomedical research and Engineering.

| Discipline | 01 vs 08 | 02 vs 08 | 03 vs 08 | 04 vs 08 | 05 vs 08 | 06 vs 08 | 07 vs 08 |
|---|---|---|---|---|---|---|---|
| Biology | 1.622 | 0.743 | 0.458 | 0.309 | 0.209 | 0.136 | 0.081 |
| Biomedical research | 1.585 | 0.714 | 0.407 | 0.279 | 0.190 | 0.132 | 0.080 |
| Chemistry | 1.814 | 0.825 | 0.542 | 0.376 | 0.257 | 0.162 | 0.104 |
| Clinical medicine | 1.638 | 0.796 | 0.463 | 0.309 | 0.204 | 0.143 | 0.090 |
| Earth and space science | 1.320 | 0.816 | 0.478 | 0.312 | 0.223 | 0.148 | 0.091 |
| Engineering | 1.089 | 0.651 | 0.404 | 0.290 | 0.194 | 0.135 | 0.079 |
| Mathematics | 1.382 | 0.925 | 0.611 | 0.419 | 0.306 | 0.235 | 0.171 |
| Physics | 1.312 | 0.590 | 0.393 | 0.274 | 0.185 | 0.125 | 0.072 |

*Table 4: Comparisons of average differences in classification of publications between any given year*

---

[5] For example, if a publication placed in the third quartile in 2003, but by 2008 placed in the first quartile; and another publication the other way around, the average shift is (2+2)/2.

*and the final year (2008)*

Next we describe the trends in average variation of class for three disciplines (Figure 6), one from each of the three clusters identified in Section 3.1. For all three disciplines we see a drastic reduction in average shift between 2001 and 2002, and then a very similar linearization of the trends from the third year and onward after publication.

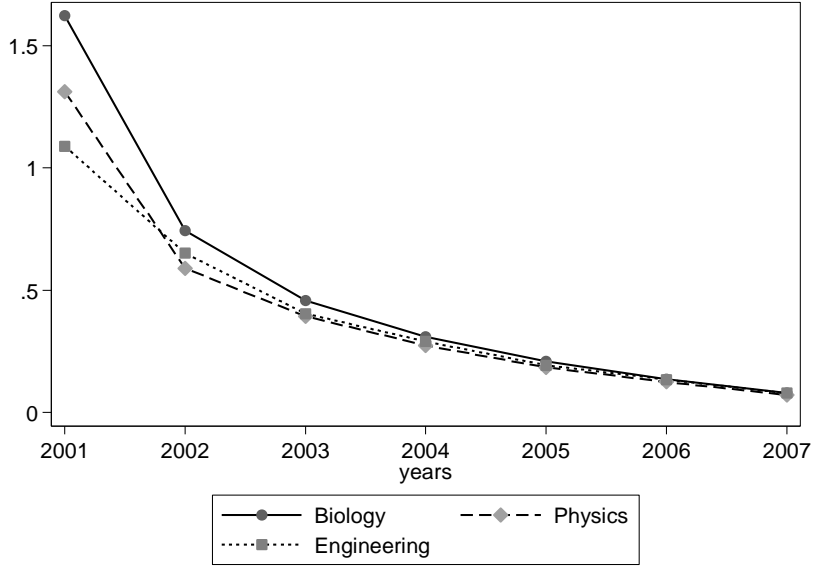

*Figure 6: Average differences of classes between any given year and the final year (2008)*

Another aspect of interest for this type of analysis concerns the phenomenon of outliers, or those publications where the evaluations conducted in a given year, compared to the benchmark, show differences of 2 or 3 classes (Table 5).

| Comparison | Class variation | Biology | Biomedical research | Chemistry | Clinical medicine | Earth and space science | Engineering | Mathematics | Physics |
|---|---|---|---|---|---|---|---|---|---|
| 01 vs 08 | ≥2 | 32.4 | 30.6 | 36.7 | 32.3 | 27.0 | 22.8 | 25.1 | 34.4 |
|  | 3 | 16.7 | 17.1 | 18.1 | 16.9 | 12.9 | 11.2 | 15.8 | 20.8 |
| 02 vs 08 | ≥2 | 12.8 | 11.9 | 14.6 | 14.0 | 16.0 | 13.1 | 16.4 | 10.1 |
|  | 3 | 4.9 | 4.7 | 4.9 | 5.5 | 6.9 | 5.6 | 9.6 | 3.3 |
| 03 vs 08 | ≥2 | 5.1 | 4.2 | 6.4 | 5.1 | 7.6 | 6.9 | 10.1 | 4.4 |
|  | 3 | 1.0 | 0.8 | 1.0 | 1.0 | 1.8 | 2.0 | 4.3 | 0.9 |
| 04 vs 08 | ≥2 | 1.8 | 1.1 | 2.4 | 2.2 | 3.2 | 4.0 | 4.9 | 2.0 |
|  | 3 | 0.1 | 0.1 | 0.2 | 0.2 | 0.3 | 0.7 | 1.9 | 0.2 |
| 05 vs 08 | ≥2 | 0.6 | 0.3 | 0.7 | 0.6 | 1.2 | 2.0 | 2.6 | 0.7 |
|  | 3 | 0.0 | 0.0 | 0.0 | 0.0 | 0.0 | 0.2 | 0.7 | 0.0 |
| 06 vs 08 | ≥2 | 0.2 | 0.0 | 0.2 | 0.1 | 0.4 | 0.9 | 0.9 | 0.2 |
|  | 3 | 0.0 | 0.0 | 0.0 | 0.0 | 0.0 | 0.0 | 0.0 | 0.0 |
| 07 vs 08 | ≥2 | 0.0 | 0.0 | 0.0 | 0.0 | 0.0 | 0.2 | 0.3 | 0.0 |
|  | 3 | 0.0 | 0.0 | 0.0 | 0.0 | 0.0 | 0.0 | 0.0 | 0.0 |

*Table 5: Percentage of publications showing by a substantial class variation between any given year and the final year (outliers)*

As in the preceding analysis, the results show that classifications as of 31/12/2001 are clearly unreliable: between a quarter and a third of publications show differences in classification equal to or more than two steps from benchmark (31/12/2008). If we analyze the differences in classification at year end 2002 compared to benchmark, we observe that the greatest differences in outliers concerns Mathematics, once again as expected. By year end 2002, the outliers with shifts of two or more classes represent 16.4% of the total publications, or 9.6% if we consider only publications with shifts of 3 classes[6]. But by year end 2003 the outliers are noticeably less, descending to less than 8%, for the shifts of two classes, and less than 2% for the shifts of three classes: the only discipline that exceeds these percentages of major shifts is Mathematics.

For the Physics discipline, evaluation of publication impact already shows a very modest number of outliers by 2002. Biomedical research shows a similar situation, and both of these disciplines have descended to nil outliers from benchmark by year end 2006.

## 4. Conclusions

In recent years we have seen increasing reliance on bibliometric indicators, both used independently and in "informed peer review" types of evaluation exercises, and generally involving indicators based on citation of publications. One of the critical issues concerning reliability of these indicators concerns the rapidity with which citations develop. Citation count can be considered a reliable proxy of real impact of a work only if observed at sufficient distance in time from the date of publication. In this work we have attempted to provide quantitative meaning to the term "sufficient", analyzing citation speeds and patterns for Italian WoS-indexed works (published 2001) under a varying citation window of up to eight years.

The results confirm previous literature indicating that different fields show different citation patterns (Garfield, 1972; Egghe and Rousseau, 1990; Moed, 2005, Althouse et al. 2009). In Biology, Biomedical research, Chemistry, Clinical medicine and Physics, the peak in citations occurs in the second year after publication, after which citations stabilize or start a decline. Citations for a second group of disciplines follow a more regular and slower-growing trend: for Earth and space science, Engineering, and especially for Mathematics, the peak of citations occurs in the last year of the time window. There does not seem to be any difference in patterns between the sub-groups of highly and poorly cited publications within each discipline, apart from the obvious differences in absolute intensity.

The citation speed is also quite different for clusters of disciplines. For Biology, Biomedical research, Chemistry, Clinical medicine and Physics, over 85% of the articles that are cited by 2008 have already received first citation within the second year following publication. For Earth and space science, Engineering and Mathematics, the analogous values are respectively 77.6%, 73.8% and 64.8% of the final articles cited. Mathematics seems to be an outlier in this analysis, in the sense that papers in this field collect citations very slowly. This may be due to consolidated practices in this discipline. The average number of citations per article is relatively small and only articles whose findings are significantly used are cited. Furthermore, results are often appreciated several years after publication. Apart from Mathematics then, given the data, we argue that the accuracy in measurement of publication impact falls within acceptable levels for citation windows of two years for the first cluster of disciplines noted and three years for the second group. In other words, a time lapse of two or three years between date of publication and citation observation appears a sufficient guarantee of robustness in impact indicators. A greater time lag would offer greater accuracy, though with ever decreasing incremental effect, but would also add further delay in carrying out the evaluation. The appropriate choice of citation window is a compromise between accuracy and timeliness in measurement. This trade-off should be carefully examined given the specific objectives and context of the exercise. We note that the estimate of accuracy, which here we conducted for only one year of publications, would necessarily be averaged over all the years to be included in a true evaluation exercise (typically three or more). Finally, we note that the ultimate objective of evaluation exercises is not just to measure the impact of the publications, but also the scientific productivity of individuals and organizations. Thus it would be very pertinent to investigate the effects of the citation window length on the rankings of individuals and organizations by scientific productivity: this is a study that the authors are currently beginning.

---

[6] Note that a difference of 3 indicates that the publication shifted from class 1 to 4 (or the opposite), or moved from no citations to a number sufficient for class 2.